# Observation of Floquet-Bloch states on the surface of a topological insulator

Y. H. Wang†, H. Steinberg, P. Jarillo-Herrero & N. Gedik*

Department of Physics, Massachusetts Institute of Technology, Cambridge MA 02139.

* To whom correspondence and requests for materials should be addressed. Email: gedik@mit.edu

† Present address: Department of Physics and Applied Physics, Stanford University, Stanford, CA 94305

**The unique electronic properties of the surface electrons in a topological insulator are protected by time-reversal symmetry. Circularly polarized light naturally breaks time-reversal symmetry, which may lead to an exotic surface quantum Hall state. Using time- and angle-resolved photoemission spectroscopy, we show that an intense ultrashort mid-infrared pulse with energy below the bulk band gap hybridizes with the surface Dirac fermions of a topological insulator to form Floquet-Bloch bands. These photon dressed surface bands exhibit polarization-dependent band gaps at avoided crossings. Circularly polarized photons induce an additional gap at the Dirac point, which is a signature of broken time-reversal symmetry on the surface. These observations establish the Floquet-Bloch bands in solids and pave the way for optical manipulation of topological quantum states of matter.**

Three dimensional topological insulators (TIs) host an exotic surface state that obeys the Dirac equation and exhibit spin momentum locking. The gapless surface states are protected by time-reversal symmetry (TRS), the breaking of which is predicted to lead to many exotic states (*1-3*). Doping TIs with magnetic impurities breaks TRS on the surface (*4-7*), but it also introduces disorder (*8*); the coherent interaction between light and matter is a promising alternative route towards such a broken symmetry phase (*9-12*). This coherent effect is seen in atoms and molecules as hybridized states distinctive in their energy spectra (*13, 14*) and in photonic systems as Floquet states (*15*). In solid state systems the photon-dressed bands lead to a periodic band structure in both energy and momentum called Floquet-Bloch states (*16*). In the case of TIs, an additional effect is expected to take place when circularly polarized light is coupled with the surface states: TRS will be spontaneously broken and the surface Dirac cone becomes gapped (*9, 17*).

Floquet theorem states that a Hamiltonian periodic in time has quasi-static eigenstates that are evenly spaced by the drive photon energy (*18*). These so-called Floquet states can be regarded as a time analog of Bloch states, which are the eigenstates of a Hamiltonian periodic in space (*19*). Combining the two situations, a periodic excitation on a crystalline lattice induces Floquet-Bloch bands that repeat in both momentum and energy. Just as different Bloch bands hybridize and develop band gaps at the crossing points (*19*), the crossing points between different orders ($n$) of the Floquet-Bloch bands open dynamic gaps (*16, 20*). Although conventional optical spectroscopy (*21-23*) and photoemission spectroscopy (*24, 25*) have shown many coherent phenomena under intense radiation, a technique that can probe the states in an energy-momentum resolved manner while applying a strong photoexcitation is needed to image the Floquet-Bloch states and the photo-induced band gaps on the surface of a TI.

Time- and- angle-resolved photoemission spectroscopy (TrARPES) is a powerful technique capable of resolving such photo-induced band gaps (*26-28*). When the photon energy of the laser that excites the sample is below the bulk band gap (typically less than 300 $\mathrm{meV}$), the coherent interaction between light and the TI surface states are the dominant effect (*10, 11*).

To achieve this regime, we used polarized photons at mid- infrared (MIR) wavelengths to investigate the photon-dressed surface states in TIs.

We use a time-of-flight spectrometer to simultaneously acquire the complete surface band structure of $Bi_2Se_3$ without changing the sample or the detector orientation in a pump-probe scheme (*28, 29*). The MIR pump pulses are generated from a commercial optical parametric amplifier pumped by a Titanium:Sapphire amplifier. The output pulses are tunable in wavelength from 4 μm to 17 μm with 1 μJ peak energy. The MIR pulses are focused to a 300 μm diameter spot (FWHM) on the single crystal $Bi_2Se_3$ sample at an angle of 45 degrees. The pulsewidth is estimated to be 250 fs (FWHM) from the rising edge of the momentum integrated time spectrum [Fig. S1D]. We estimate the amplitude of the electric field to be $E_0 = 2.5(\pm 1) \times 10^7\ \mathrm{V/m}$ on the surface of the sample after taking into account the loss through the optics, the geometric effect and the Fresnel reflection (*30*) at the vacuum-sample interface (*31*). The polarization of the MIR pulse is adjusted by a quarter wave-plate. Figure 1A illustrates the p-polarized case in which the electric field of light is in the plane of incidence. The component of the surface electron momentum in the plane of incidence is defined as $k_x$.

Figure 1, B-F shows energy-momentum spectra of $Bi_2Se_3$ obtained at several time delays $t$ after the intense linearly polarized MIR excitation. At $t = -500$ fs, the probe UV pulse is ahead of the MIR pulse and the band structure is similar to that of an unperturbed system [Fig. 1B] (*28*). The asymmetry in the spectral intensity about zero momentum $\Gamma$ is due to the coupling between the linearly polarized UV pulse and the spin-orbit texture of the surface states of TI (*29*). When the two pulses start to overlap in time ($t = -200$ fs), replicas of the surface Dirac cone appear above and below the original band [Fig. 1C]. The energy difference between the bands equals exactly the pump photon energy of $\hbar\omega = 120$ meV, where $\hbar$ is the reduced Planck constant and $\omega$ is the angular frequency of light. The intensity of such side bands becomes stronger while the original band gets weaker when the pump and probe pulses are completely overlapped at $t = 0$ [Fig. 1D]. When the side bands become weaker again at 200 fs [Fig. 1E], the spectral intensity still above the Fermi level occupies the higher energy states of the original surface and bulk bands rather than the sidebands. This suggests the onset of a thermally excited carrier population as has been observed previously with 1.5 eV photoexcitations (*26, 28*). This excited population is still present at 500 fs [Fig. 1F], consistent with the observed photoexcited carrier cooling time in this material (*26, 28*). The fact that the sidebands are only present during the time duration of the pump pulse suggests that it is a consequence of the coherent interaction between the MIR photons and the electron system. In a TrARPES experiment, strong IR field may also generate band replicas through laser assisted photoemission (LAPE) (*24, 25*). Because this effect is caused by the absorption of MIR photons by the near-free electrons in the final states of photoemission, the LAPE bands do not open band gaps at where they cross and their intensity is at a minimum in the direction perpendicular to the light polarization, which is inconsistent with our findings (*31*).

To see how the photon-dressed bands change with respect to the polarization of light and whether there are band gaps, we compare the energy momentum spectra taken at $t = 0$ fs along the directions parallel and perpendicular to the electric field of the linearly polarized pump [Fig. 2]. Compared with the sidebands along $k_x$ [Fig. 2A], the sidebands in the $k_y$ direction shows a higher order ($n = +2$) more clearly [Fig. 2B]. More importantly the bands are much stronger and show features which deviate from a mere stacking of Dirac cones at energies close to the $n = 0$ Dirac point ($0.15\text{eV} < E < 0.3\text{eV}$). The otherwise linearly dispersing $n = 0$ cone becomes discontinuous and distorted at $E = 0.1\text{eV}$ and $0.22\text{eV}$ [Fig. 2B, pink triangles].

In order to better visualize these features, we subtract the spectrum taken at $t = -500$ fs from the spectrum at $t = 0$ fs, which includes the spectral contribution from the unperturbed surface bands due to the finite pulse width (*31*). The difference spectrum along $k_y$ [Fig. 2D] shows a pattern resembling an '∞' sign centered at the $n = 0$ Dirac point. There is a replica of this ∞ pattern at one pump photon energy (120 meV) above the one around $n = 0$ Dirac point, suggesting that this is a coherent feature on the photo-induced bands. $E = 0.1\text{eV}$ is the middle point of these two ∞, suggesting that the kinks in the raw spectrum [Fig. 2B] correspond to opening of band gaps. These features are in contrast to the difference spectrum in the $k_x$ direction, where the pattern is simply composed of three Dirac cones shifted by $\hbar\omega$ [Fig. 2C]. This momentum-dependent feature is consistent with the prediction of photon-dressed Dirac bands in graphene under linearly polarized light (*32-34*), where Floquet-Bloch bands open band gaps along the direction perpendicular to the electric field whereas the bands with momentum parallel to the field remain gapless.

For an ideal Dirac cone [Fig. 2E], the avoided crossing between adjacent orders are centered at momenta $\pm\omega/2v$, where $v$ is the Fermi velocity of the surface states. In our case, we can see in the difference spectra the crossing of bands at $k_x \sim \pm 0.03$ Å$^{-1}$ [Fig. 2C] while the band gaps appear at similar values of $k_y$ [Fig. 2D]. This is slightly bigger than the expected value of $\omega/2v = 0.02$ Å$^{-1}$ (using $\hbar\omega = 120$ meV and $\hbar v = 3$ eVÅ [Fig. 1A]), likely resulting from the reduced velocity of bands below the Dirac point [Fig. 1]. At $k_y = \pm 0.06$ Å$^{-1}$ where $n = \pm 1$ cross, there is no resolvable gap [Fig. 2D]. Therefore, the "∞" pattern around the $n = 0$ Dirac point is a result of the crossings and avoided crossings between $n = 0$ and $n = \pm 1$ [Figs. 2, D and E]. The energy distribution curve [Fig. S1A] through the avoided crossing shows a $2\Delta$ value of 62 meV. Along with the crossings and avoided crossings between different orders, the gap size is consistent with the theory of photon-dressed Dirac systems based on the Floquet picture (*31-34*).

Having shown that linearly polarized MIR photons couple with the surface states into Floquet-Bloch bands, we investigate how these bands change when the photons become circularly polarized. Unlike the bands under linearly polarized light, the spectrum under circularly-polarized light show avoided crossing around $\pm 0.03$ Å$^{-1}$ along both $k_x$ and $k_y$ [Figs. 3, A-D]. Spectral cuts along other in-plane directions through $\Gamma$ [Fig. S3C] further show that the dynamical gap does not close for any direction. This is consistent with the Floquet-Bloch bands generated by the coupling between circularly polarized in-plane electric field and the surface states [Fig. 3E].

Knowing that the surface electrons and the rotating electric field of the circularly polarized MIR pulse are coupled, we examine the Dirac point of the induced Floquet-Bloch bands for signatures of TRS breaking [Fig. 3F]. The spectral weight above and below the Dirac point increases at $t = 0$ for both polarizations [Figs. 2, C and D and 3, C and D]. To be able to resolve any gap around the Dirac point, we convert the intensity spectra for both polarizations [Figs. 2B and 3B] into energy distribution curves (EDC) by integrating the spectra over a small momentum window (0.005$\text{Å}^{-1}$) at each momentum point along $k_y$ [Fig. 4]. The EDCs show that only the spectrum obtained under circularly polarized excitations opens a band gap at the Dirac point [Fig. 4A], whereas such a gap is absent in the curves taken under linearly polarized light [Fig. 4B]. The EDC through the $\Gamma$ point under circularly polarized light clearly show two peaks at the Dirac point [Fig. 4C blue]. (The small hump in the EDC under linearly polarized excitation [Fig. 4C red] is a result of the finite momentum resolution of our spectrometer (~0.005$\text{Å}^{-1}$) and is also visible in the EDC of unperturbed spectrum [Fig. 4C black].) This result is similarly observed in the difference spectrum between the two polarizations as negative spectral weight at the Dirac point [Fig. S4A]. By fitting the energy distribution curve through the $\Gamma$ point in Figure 4A, we obtain the band gap $2\kappa = 53\ meV$ [Fig. S1B]. Using the $2\Delta$ value we obtained from the dynamic gap (*31*), we find that the Dirac gap is consistent with the gap resulting from the two-photon process (absorbing and emitting a photon) that is only allowed under circular polarization that breaks TRS (*9, 34*).

The surface Dirac Hamiltonian hybridized with circularly polarized light essentially result in a Chern insulator as originally proposed by Haldane (*35*). In his model, alternating current loops break TRS locally while the total magnetization remains zero. Whereas such microscopic periodic current configuration is difficult to implement, the Hamiltonian of graphene per valley in the Haldane model is basically the same as that of a circularly polarized electromagnetic field coupled to a Dirac band (*9*), which is realized in our experiment. These observations suggest the existence of a photoinduced anomalous quantum Hall phase without Landau levels where topological magnetoelectric effect described by axion electrodynamics may exist (*5*).

The quantitative agreement between the measured gap sizes and the models based on Floquet theory shows the general applicability of our result to Dirac systems. These observations open up an avenue for optical controlling and switching topological orders and may provide a platform for realizing Floquet Majorana modes in these materials (*1,2*).

## References and Notes

(1) M. Z. Hasan and C. L. Kane, "Topological insulators," *Rev. Mod. Phys.* **82**, 3045, (2010).

(2) X. L. Qi and S. C. Zhang, "Topological insulators and superconductors," *Rev. Mod. Phys.* **83,** 1057 (2011)

(3) J. E. Moore, "The birth of topological insulators," *Nature,* **464**, 194, (2010).

(4) C. Z. Chang, *et al.*, "Experimental Observation of the Quantum Anomalous Hall Effect in a Magnetic Topological Insulator," *Science,* **340**, 167, (2013).

(5) X. L. Qi, T. L. Hughs, and S. C. Zhang, "Topological field theory of time-reversal invariant insulators," *Phys. Rev. B,* **78**, 195424, (2008).

(6) Y. L. Chen, *et al.*, "Massive Dirac Fermion on the Surface of a Magnetically Doped Topological Insulator," *Science,* **329**, 659, (2010).

(7) S. Y. Xu, *et al.*, "Hedgehog spin texture and Berry's phase tuning in a magnetic topological insulator," *Nat. Phys.,* **8**, 616, (2012).

(8) H. Beidenkopf, *et al.*, "Spatial fluctuations of helical Dirac fermions on the surface of topological insulators," *Nat. Phys.,* **7**, 939, (2012).

(9) T. Kitagawa, T. Oka, A. Brataas, L. Fu, and E. Demler, "Transport properties of nonequilibrium systems under the application of light: Photoinduced quantum Hall insulators without Landau levels," *Phys. Rev. B,* **84**, 235108, (2011).

(10) N. H. Lindner, G. Rafael and V. Galitski, "Floquet topological insulator in semiconductor quantum wells," *Nat. Phys.,* **7**, 490, (2011).

(11) J. I. Inoue and A. Tanaka, "Photoinduced Transition between Conventional and Topological Insulators in Two-Dimensional Electronic Systems," *Phys. Rev. Lett.,* **105**, 017401, (2010).

(12) O. V. Yazyev, J. E. Moore and S. G. Louie, "Spin Polarization and Transport of Surface States in the Topological Insulators $Bi_2Se_3$ and $Bi_2Te_3$ from First Principles," *Phys. Rev. Lett.,* **105**, 266806, (2010).

(13) C. Cohen-Tannoudji, J. Rupont-Roc and G. Grynberg, "Atom-Photon Interaction", Wiley, 1992.

(14) E. Goulielmakis, *et al.*, "Real-time observation of valence electron motion," *Nature,* **466**, 739, (2010).

(15) M. Rechstman, *et al.*, "Photonic Floquet topological insulators," *Nature,* **496**, 196, (2013).

(16) F. M. Faisal and J. Z. Kaminski, "Floquet-Bloch theory of high-harmonic generation in periodic structures," *Phys. Rev. A,* **56**, 748, (1997).

(17) B. Fregoso, Y. H. Wang, N. Gedik and V. Galitski, "Driven Electronic States at the Surface of a Topological Insulator," *arXiv:1305.4145.*

(18) V. M. Galitskii, S. P. Goreslavskii and V. F. Elesin, "Electric and Magnetic Properties of a Semiconductor in the Field of a Strong Electromagnetic Wave," *Sov. Phys. JETP,* **30**, 117, (1970).

(19) C. Kittel, "Introduction to Solid State Physics," Wiley, 2004.

(20) H. Sambe, "Steady States and Quasienergies of a Quantum-Mechanical System in an Oscillating Field," *Phys. Rev. A,* **7**, 2203, (1976).

(21) M. Schultze, *et al.*, "Controlling dielectrics with the electric field of light," *Nature,* **493**, 75, (2013).

(22) Q. T. Vu, *et al.*, "Light-Induced Gaps in Semiconductor Band-to-Band Transitions," *Phys. Rev. Lett.,* **92**, 217403, (2004).

(23) S. Ghimire, *et al.*, *Nat. Phys.,* "Observation of high-order harmonic generation in a bulk crystal," **7**, 138, (2011).

(24) G. Saathoff, L. Miaja-Avila, M. Aeschlimann, M. M. Murnane, and H. C. Kapteyn, "Laser-assisted photoemission from surfaces," *Phys. Rev. A,* **77,** 022903, (2008).

(25) L. Miaja-Avila, *et al.*, "Ultrafast studies of electronic processes at surfaces using the laser-assisted photoelectric effect with long-wavelength dressing light," *Phys. Rev. A,* **79**, 030901, (2009).

(26) J. Sobota, *et al.*, "Ultrafast Optical Excitation of a Persistent Surface-State Population in the Topological Insulator $Bi_2Se_3$," *Phys. Rev. Lett.,* **108**, 117403, (2012).

(27) M. Hajlaoui, *et al.,* "Ultrafast Surface Carrier Dynamics in the Topological Insulator $Bi_2Te_3$," *Nano. Lett.,* **12**, 3532, (2012).

(28) Y. H. Wang, *et al.*, "Measurement of Intrinsic Dirac Fermion Cooling on the Surface of the Topological Insulator $Bi_2Se_3$ Using Time-Resolved and Angle-Resolved Photoemission Spectroscopy," *Phys. Rev. Lett.,* **109**, 127401, (2012).

(29) Y. H. Wang, *et al.*, "Observation of a Warped Helical Spin Texture in $Bi_2Se_3$ from Circular Dichroism Angle-Resolved Photoemission Spectroscopy," *Phys. Rev. Lett.,* **107,** 207602, (2011).

(30) A. D. LaForge, *et al.*, "Optical characterization of Bi2Se3 in a magnetic field: Infrared evidence for magnetoelectric coupling in a topological insulator material," *Phys. Rev. B,* **81**, 125120, (2010).

(31) "Materials and methods are available as supporting material on Science Online".

(32) S. V. Syzranov, M. V. Fistul, and K. B. Efetov, "Effect of radiation on transport in graphene," *Phys. Rev. B.,* **78**, 045407, (2008).

(33) Y. Zhou and M. W. Wu, "Optical response of graphene under intense terahertz fields," *Phys. Rev. B.,* **83**, 245436, (2011).

(34) T. Oka and H. Aoki, "Photovoltaic Hall effect in graphene," *Phys. Rev. B,* **79**, 081406, (2009).

(35) F. D. Haldane, "Model for a Quantum Hall Effect without Landau Levels: Condensed-Matter Realization of the "Parity Anomaly"," *Phys. Rev. Lett.,* **61**, 2015, (1988).

## Acknowledgments:

This work is supported by Department of Energy award numbers DE-FG02-08ER46521and DE-SC0006423 (data acquisition and analysis), Army Research Office (ARO-DURIP) Grant No. W911NF-09-1-0170 (ARTOF spectrometer). HS and PJH have been supported by the US DOE, BES Office, Division of Materials Sciences and Engineering under Award DE-SC0006418 (materials growth). This work made use of the MRSEC Shared Experimental Facilities supported by NSF under award No. DMR-0819762. We are grateful for the stimulating discussions with L. Fu, T. Kitagawa, T. Oka, B. Fregoso and V. Galitski.

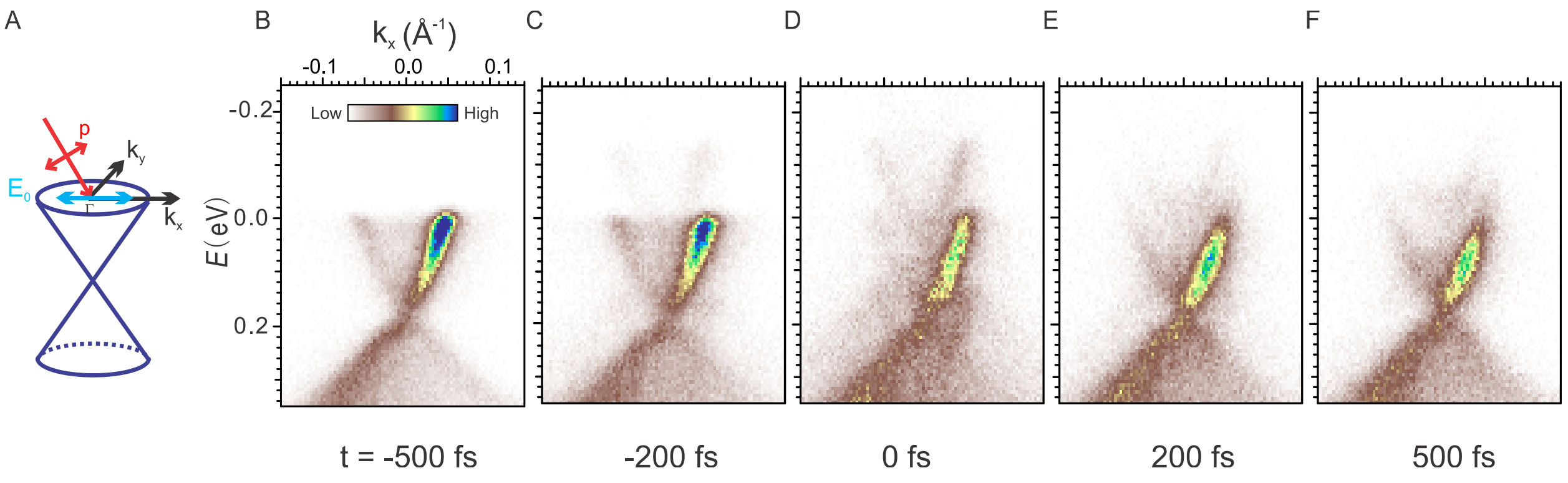

**Fig. 1. Angle resolved photoemission spectra (APRES) of $Bi_2Se_3$**. (**A**) A sketch of the experimental geometry for the p-polarized case. $k_x$ is defined to be the in-plane electron momentum parallel to the pump scattering plane**.** (**B-F**) ARPES data for several pump-probe time delays $t$ (values indicated in the figure) under strong linearly polarized mid-infrared (MIR) excitation of wavelength $\lambda = 10$ μm.

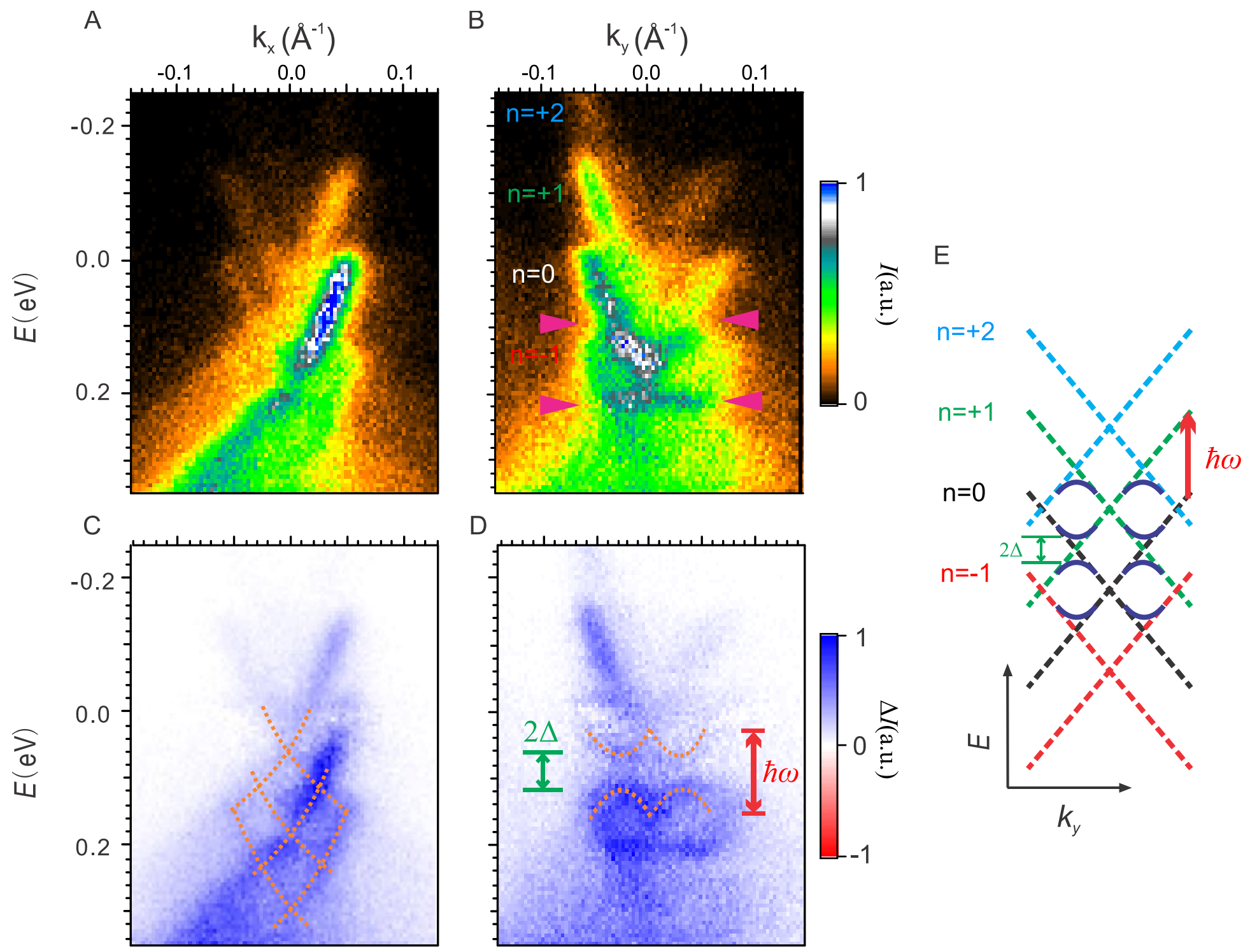

**Fig. 2. ARPES spectra at 0 fs time delay under linearly polarized MIR excitation**. (**A**) and (**B**) show the energy-momentum spectra through Γ along $\boldsymbol{k_x}$ and $\boldsymbol{k_y}$ direction respectively; (**C**) and (**D**) are the same spectra after subtracting the spectra at $\boldsymbol{t} = -\mathbf{500\ fs}$ (*35*). Dashed orange lines are guides to the eye. (**E**) Sketch of the sidebands of different order as induced by the mid infrared excitation. Avoided crossing occurs along $\boldsymbol{k_y}$, leading to a pattern of ∞ around the Dirac point (see text). $\boldsymbol{\omega}$ is the drive photon frequency and $\boldsymbol{\Delta}$ is the half gap at the avoided crossings.

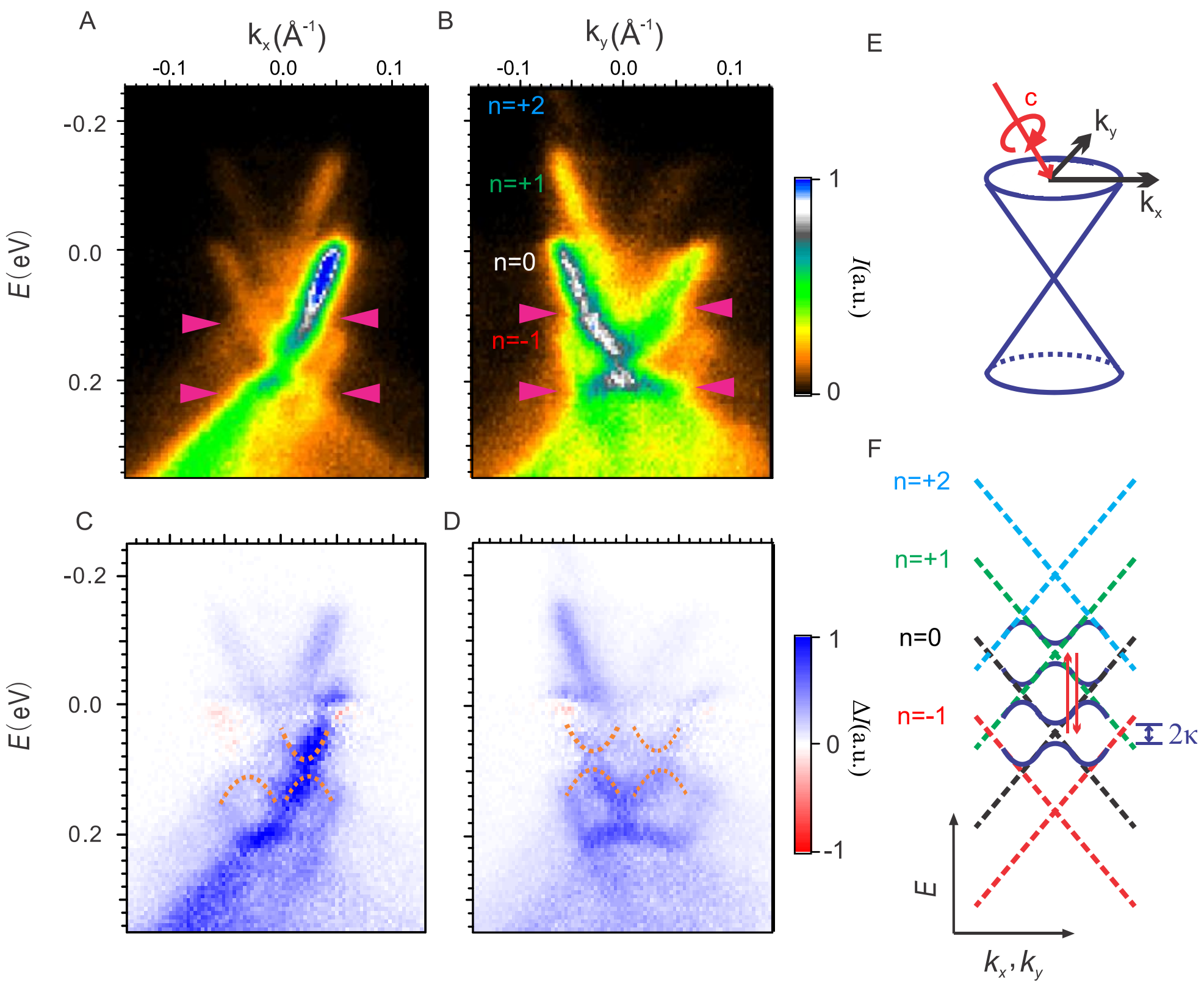

**Fig. 3. ARPES spectra at 0 fs time delay under circularly polarized MIR excitation** Panels (**A-D**) are analogous to the corresponding panels in Figure 2. (**E**) The projection of the electric field on the surface plane (light blue) is elliptical. And the avoided crossing appears along both directions. The chirality allows emitting and absorbing a photon to open a band gap at the Dirac point as sketched in the sideband diagram in (**F**). **κ** is half of the band gap at the Dirac point.

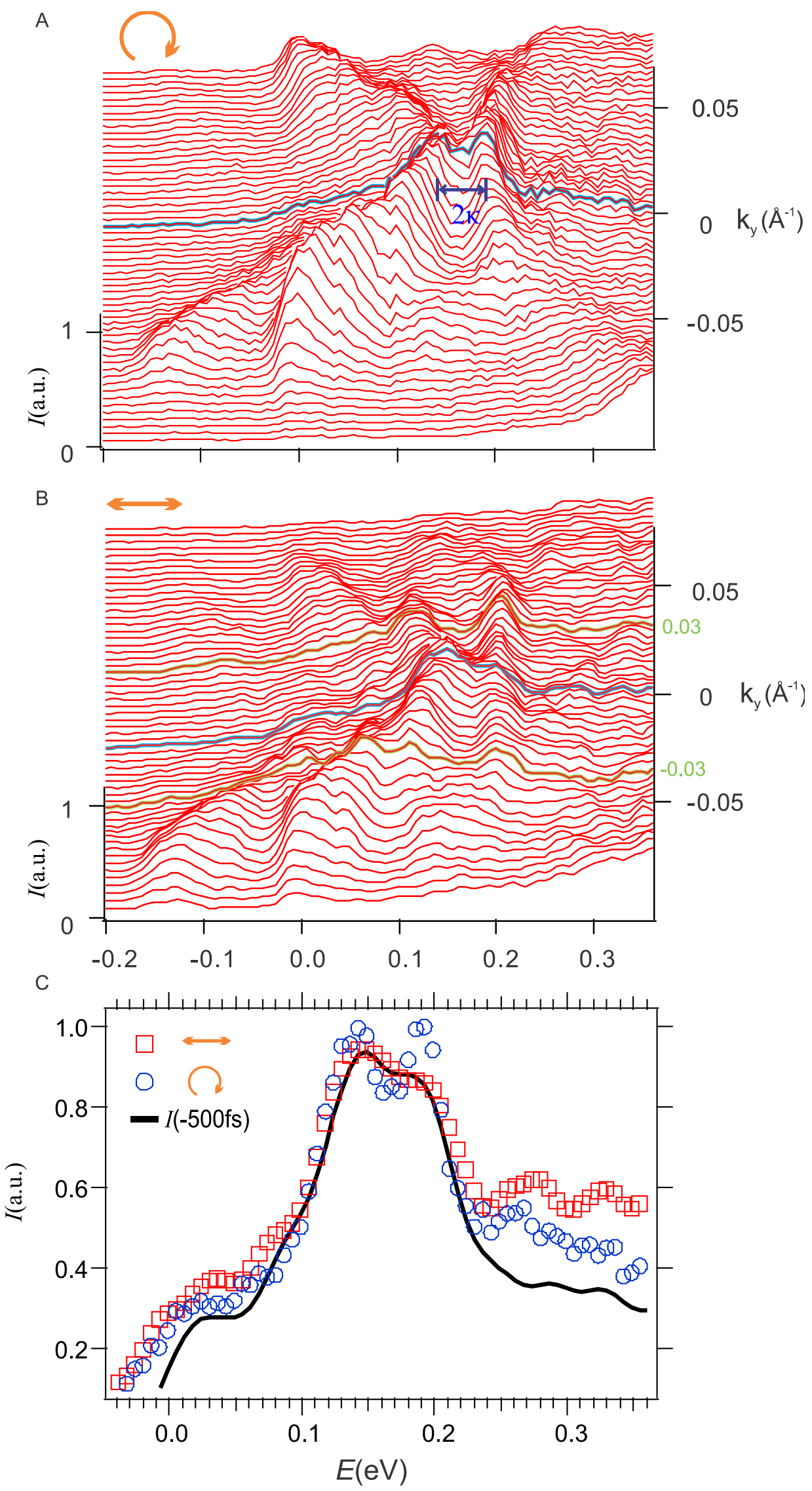

**Fig. 4. Band gap at the Dirac point induced by circularly polarized MIR pulse**. (**A**) and **(B)** show the energy distribution curves along $\boldsymbol{k_y}$ obtained from intensity spectra in Fig. 2(B) and Fig. 3(B) under circularly and linearly polarized MIR excitation respectively. The blue curves are the ones passing through the $\boldsymbol{\Gamma}$ point. The green curves in (**B**) are through the avoided-crossing momenta. (**C**) EDCs through $\boldsymbol{\Gamma}$ under linearly (red squares) and circularly (blue circles) polarized pump at $\boldsymbol{t = 0\ \mathrm{fs}}$ as well as the EDC through $\boldsymbol{\Gamma}$ taken at $\boldsymbol{t = -500\ \mathrm{fs}}$ (black)

# Supplementary Materials

## Materials and Methods

$Bi_2Se_3$ single crystals were prepared by melting Bi and Se with 2:4.06 ratio in a quartz tube and heating to 850 ℃ for 2 days. The melt was subsequently cooled to 550 ℃ over 5 days, annealed at the same temperature for 5 more days and cooled to room temperature. At low temperatures, the material exhibits bulk electron densities of about $2 \times 10^{17}\mathrm{cm}^3$.

## Supplementary Text

Fitting energy distribution curves

In this section, we first show the fitting procedure of the energy distribution curves (EDC) to obtain the gap sizes $2\Delta$ and $2\kappa$ [Fig. S1]. (The factor of 2 is used here following the convention in the literature.) We use multi-peak Voigt functions with a base line to fit these EDCs. A Voigt function is a convolution of a Lorentz function, which represents the intrinsic linewidth of the spectrum, and a Gaussian function which represents the instrumental resolution and is fixed at 20 meV in our fittings. Since we are only interested in the distance between the peaks, the choice of the resolution does not impact the gap size in a noticeable way. The baseline is a polynomial function to account for the bulk band intensity which has a broad energy spread outside the window of interest ($\mathrm{E} > 0.25\mathrm{eV}$).

For the EDC through $k_y = 0.03\ \text{Å}^{-1}$ [Fig. S1A] where the avoided crossing is observed [Fig. 2D], we fit it with four Voigt functions. These peaks are from the upper and lower branch of the Dirac cone on the $n = 0$ and $n = 1$ order [Fig. 2E]. We obtain $2\Delta = 62 \pm 5\ \mathrm{meV}$ from the distance between the second and third peak as they represent the anti-crossing between $n = 0$ and $n = 1$. For the EDC through Γ under circularly polarized light [Fig. S1B], we use two peaks for the Dirac point of the $n = 0$ order because peaks for other orders are much smaller. We obtain the gap size $2\kappa = 53 \pm 4\ \mathrm{meV}$.

We compare the fitted gap sizes with the expected gaps based on Floquet band theory (*17, 32-34*). We define the coupling potential $V \equiv evE_0/\omega$ and coupling constant $\beta \equiv V/(\hbar\omega)$, where $E_0$ is the amplitude of the electric field of light. Under perturbation theory with rotating wave approximation, the gap at $\pm\omega/2v$ when the light is linearly polarized is proportional to the field (*17,32,34*):

$$\mathbf{2\Delta} = \frac{\boldsymbol{evE_0}}{\boldsymbol{\omega}}\ \boldsymbol{mod}\ \boldsymbol{\hbar\omega} \qquad \text{Eq. S1}$$

Using $E_0 = 2.5 \times 10^7\ \mathrm{V/m}$ (see the detailed explanation below), we get $2\Delta = 63\ \mathrm{meV}$ and $\beta = 0.53$ .

Now we calculate the expected gap at the Dirac point and compare it with the fitted value. The gap $2\kappa$ at the Dirac point is given by [*17,34*]:

$$\mathbf{2\kappa} = \sqrt{\mathbf{4V^2} + (\boldsymbol{\hbar\omega})^2} - \boldsymbol{\hbar\omega} \qquad \text{Eq. S2}$$

Note that, in the small coupling limit, Eq. S2 becomes $2\kappa = \frac{2\mathrm{V}^2}{\hbar\omega}$, which shows that the gap at the Dirac point is due to two-photon process [Fig. 3F]. Substituting $V = 63$ meV into Eq. S2, we get $2\kappa = 54$ meV. The agreement between this value and the fitted value suggests the self-consistency of our model.

We estimate the value of the electric field ($E_0 = 2.5(\pm 1) \times 10^7$ V/m ) after considering the loss at the optics, beam profile and Fresnel reflection at the sample-vacuum interface [Fig. S3D]. The measured pulse energy is 1 μJ at the output of the OPA and 0.3 μJ right before the vacuum window. The loss is due to the lenses, the mirrors and the waveplate. The transmission through the ZeSe vacuum window is 95%. Our beam profile is Gaussian with a 300 μm FWHM spot size at the focus and 250 fs FWHM temporal pulsewidth. There are two commonly used ways of estimating the energy density and fluence of Gaussian beams. One can either consider that the entire energy of the beam is contained within the FWHM or alternatively, within $1/e^2$ radius of the beam. $1/e^2$ width of the beam is equal to 1.699 times FWHM value. We estimated the electric field using both of these conventions. Assuming FWHM ($1/e^2$ width) as the beam size, we obtain an energy density of $5.38 \times 10^4$ J/m$^3$ ($1.10 \times 10^4$ J/m$^3$) incident on the sample. To calculate the value of the electric field at the sample surface, one also needs to take into account the Fresnel reflection. Since the component of the electric field that is parallel to the sample surface is continuous at the vacuum-sample interface, one can just calculate the transmitted field and its parallel component. Reflection coefficient is 44% for 10 μm (*30*) at normal incidence. Assuming that the absorption is small at our wavelengths, we obtain an index of refraction of n=4.93. Using Fresnel equations at an incidence angle of 45 degrees, we obtain a transmission coefficient of 0.32 for the electric field. This corresponds to an electric field value of $E_0 = 3.42 \times 10^7$ V/m for the FWHM beam size and $E_0 = 1.54 \times 10^7$ V/m for the case of $1/e^2$ width. We therefore use $2.5(\pm 1) \times 10^7$ V/m as our best estimate of the electric field at the surface.

## Consideration of other possible effects and presentation of additional ARPES spectra

In this section, we consider other possible effects and present additional ARPES spectra that manifest the band gaps in the Floquet-Bloch bands described in the main text.

We first provide more evidence besides the band gaps that the observed side bands are inconsistent with Laser Assisted Photoemision (LAPE) by considering the intensity of the sidebands as a function of momentum, MIR wavelength and the time delay. 1) The intensity $A_n$ of the sideband n in LAPE is known (*24, 25*) to be:

$$\boldsymbol{A_n} \approx \boldsymbol{J_n}^2\left(\frac{\boldsymbol{p} \cdot \boldsymbol{E}}{\boldsymbol{\omega}^2}\right) \qquad \text{Eq. S3}$$

Here $J_n(x)$ is the ordinary Bessel function and $\boldsymbol{p}$ is the momentum of the electron in the final state of photoemission. In the weak field limit, $J_n(x)$ vanishes as $x \to 0$ for $n > 0$. The sideband intensity is the strongest in the plane of incidence under p-polarized light. However, as can be seen in Figure 2, the sidebands along $k_y$, which is perpendicular to the plane of incidence, are about twice as strong as those along $k_x$ [Figs. 2, A and B, n=1 bands]. Furthermore, the sidebands are stronger at states around the Dirac point (energy region between the pink arrows in Figure 2B) than those close to the Fermi level ($E$ < 100 meV), which have higher kinetic energy and larger momentum. 2) Figure S1C shows that the sidebands under $\lambda = 7.5\ \mu\text{m}$ MIR excitation is much stronger than those under $\lambda = 10\ \mu\text{m}$ at the same fluence, which is the opposite of what Eq. S3 indicates. 3) The LAPE sideband intensity as a function of delay between pump and probe should be a cross-correlation of the two pulses. This is because the final states typically show no decay dynamics, which is the reason LAPE is widely used to measure the pulsewidth (*24, 25*). This is not the case for the sidebands we observe as shown in the time-resolved sideband intensity at $E = -0.2$ eV [Fig. S1D]. We can see that the decay is longer for $t > 0$ than for $t < 0$. This suggests that the MIR pulse is inducing the excitation on the surface states that is being probed by the UV pulse.

We then proceed to show that the band gaps are intrinsic and not a trivial result of matrix element effects in photoemission. The ARPES difference spectra in Figure 2 and Figure 3 are shown without guides to the eye in Figure S2. These difference spectra are obtained by subtracting the linearly (circularly) polarized spectra $I_L(t)$ ( $I_C(t)$ ) at negative time delay from their spectra at $t = 0$ fs with a scaling factor a to minimize the contribution from unperturbed bands, i.e., $\Delta I_{L(C)} = I_{L(C)}(0) - aI_{L(C)}(-500)$. The value of a is 0.37 for circularly polarized pump and 0.25 for linearly polarized pump. The value for linearly polarized pump is smaller because its $n = 0$ band is weaker. Since $I(-500)$ has much stronger intensity close to the Fermi level than at the energies close to the Dirac point [Fig. 1 B], the subtraction leads to zero intensity at $E = 0$ eV while manifesting the photo-induced spectral change away from it. The subtraction does not lead to any distortion of the bands, nor does it artificially generate any band gaps. This is confirmed by the agreement between the difference spectra [Figs. S2, B and C] and the derivative spectra [Figs. S3, A and B] obtained by taking the derivative of raw spectra with no subtraction. The band gaps manifested by these difference and derivative spectra directly reflect the kinked dispersions in the raw spectra [Figs. 2, A and B and 3, A and B].

Next, we show additional evidence of the band gap at the Dirac point under circularly polarized light by subtracting the spectrum taken with linearly polarized light. The $n = \pm 1$ bands of $I_L(0)$ are stronger than that of $I_C(0)$ but their $n = 0$ bands are the opposite [Figs. 2, B and 3, B]. Therefore, the difference spectrum $I_C(0) - I_L(0)$ shows positive (negative) $n = 0$ ($n = \pm 1$) intensity [Fig. S4A]. The difference spectrum shows that under circularly polarized light the intensity at the $n = 0$ Dirac point is reduced while states above and below the Dirac point gain intensity [Fig. S4A], suggesting that a band gap is opened at the Dirac point. In contrast, the difference spectrum under linearly polarized light at two different times ($I_L(-500) - I_L(0)$) does not show any gap at the Dirac point [Fig. S4B]. The gap in the difference spectra [Fig. S4A] is consistent with the EDCs shown in Figure 4. The small hump in the EDC through Γ under linearly polarized excitation [Fig. 4B blue and Fig. S1C red] is a result of the finite momentum resolution of our spectrometer (~0.005$\text{Å}^{-1}$) and is also visible in the EDC of unperturbed spectrum [Fig. S1C black].

Floquet theory also predicts that band gaps between $n = 1$ and $n = -1$ bands are about half of the gap size at the $n = 0$ Dirac point for circularly polarized pump (*17, 33, 34*). That means a gap size of about 26 meV, which is close to the energy resolution of our setup (20 meV). Furthermore, these two bands are very weak and broad at where they cross [Fig. 4A], making it even more difficult to reliably exam any gaps.

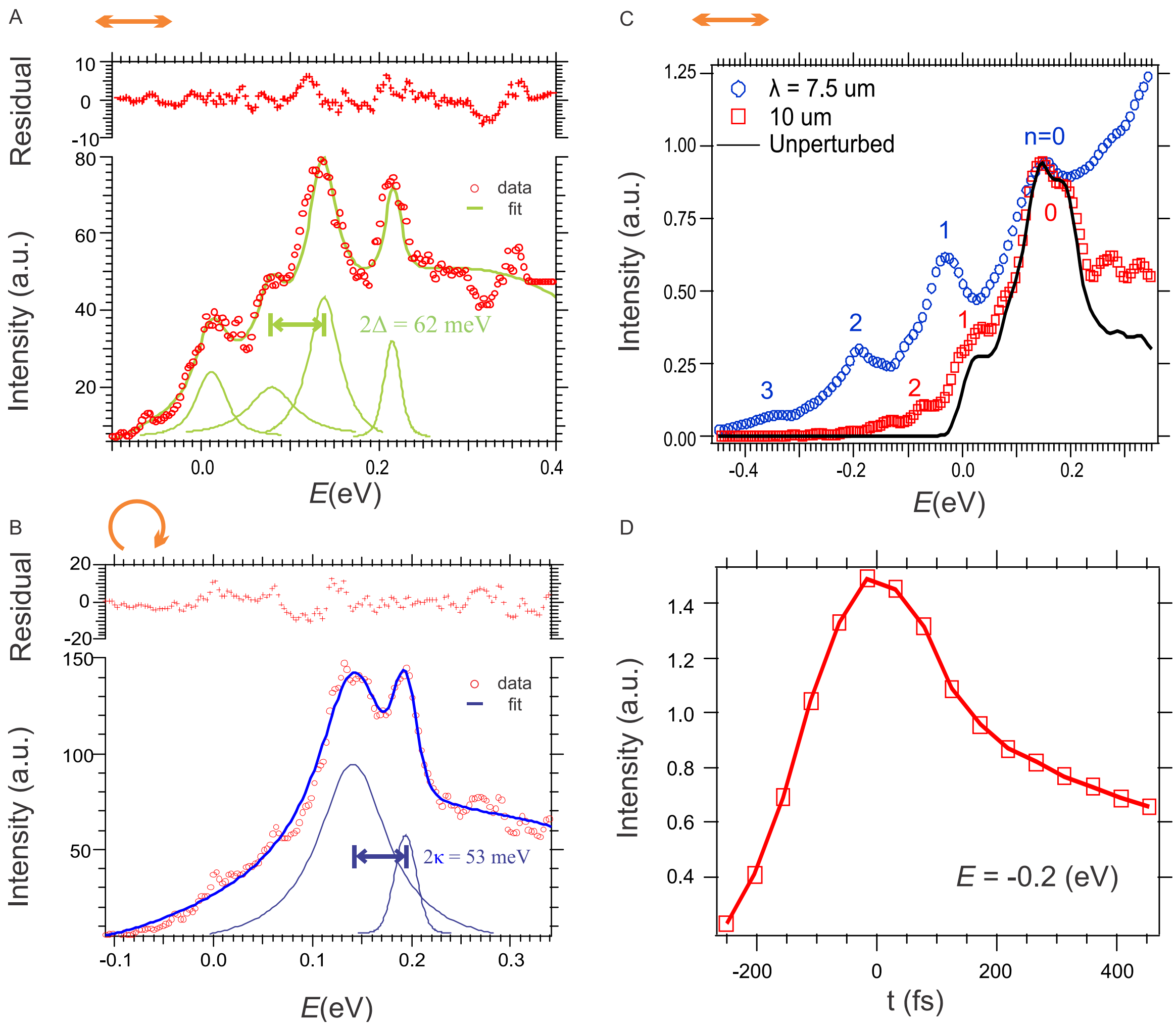

**Fig. S1.**

Fits for the energy distribution curves. **(A)** is for EDC through $k_y = 0.03$ Å$^{-1}$ obtained from the $t = 0$ fs spectra under linearly polarized light. **(B)** shows the data and fit of the energy distribution curve (EDC) through Γ under circularly polarized light. **(C)** shows EDCs through Γ under linearly polarized light at two different MIR wavelength $\lambda = 8$ μm (blue) and $\lambda = 10$ μm (red) with the same fluence. Black is the EDC taken at $t = -1$ ps under $\lambda = 10$μm. The curves are normalized to the intensity of the $n = 0$ Dirac point. **(D)** is momentum integrated spectra at $E = -0.2$ eV as a function of t.

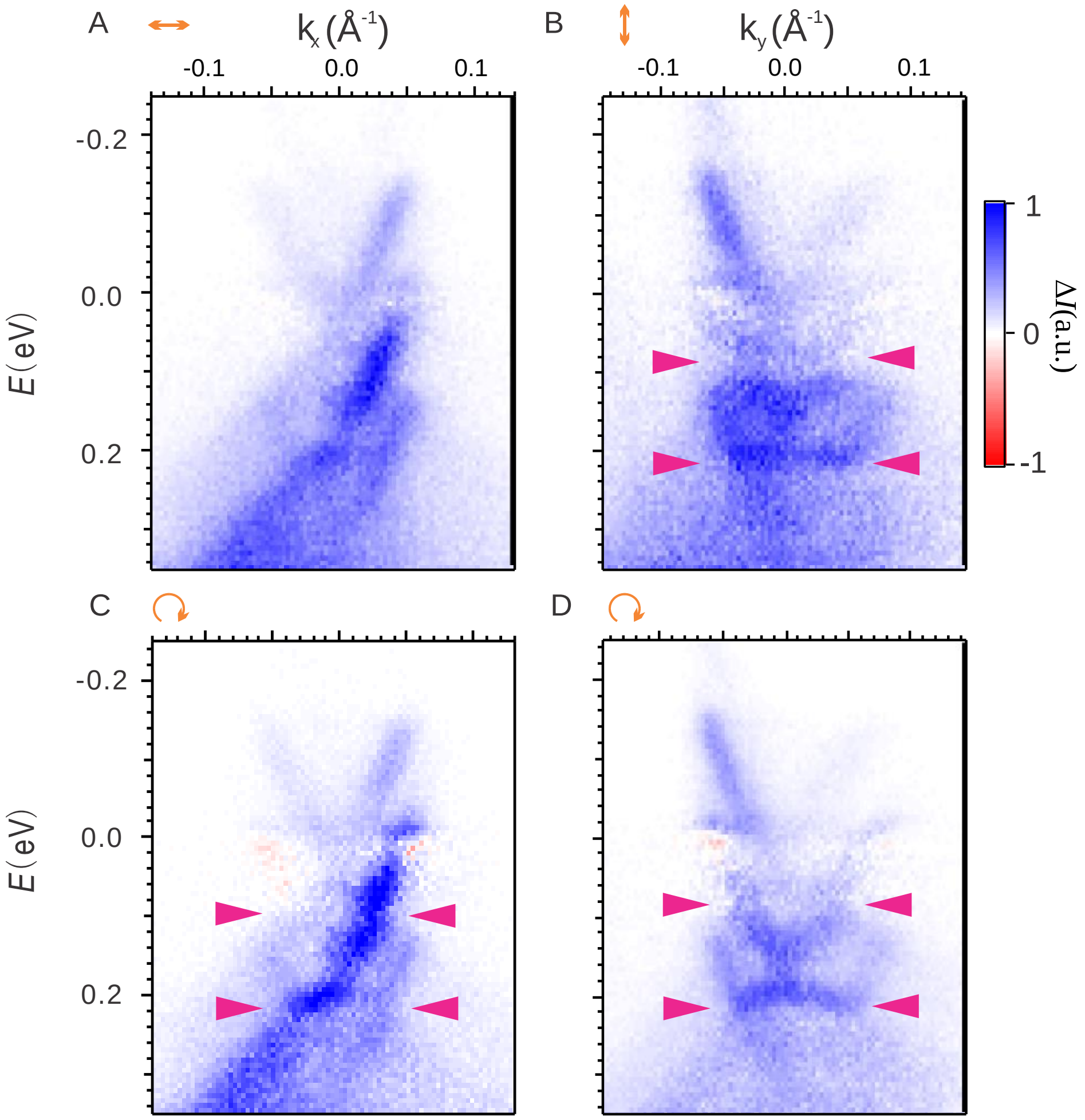

**Fig. S2**

ARPES difference spectra without guides to the eye. **(A)** and **(B)** are from Figures 2, C and D; **(C)** and **(D)** are from Figures 3, C and D.

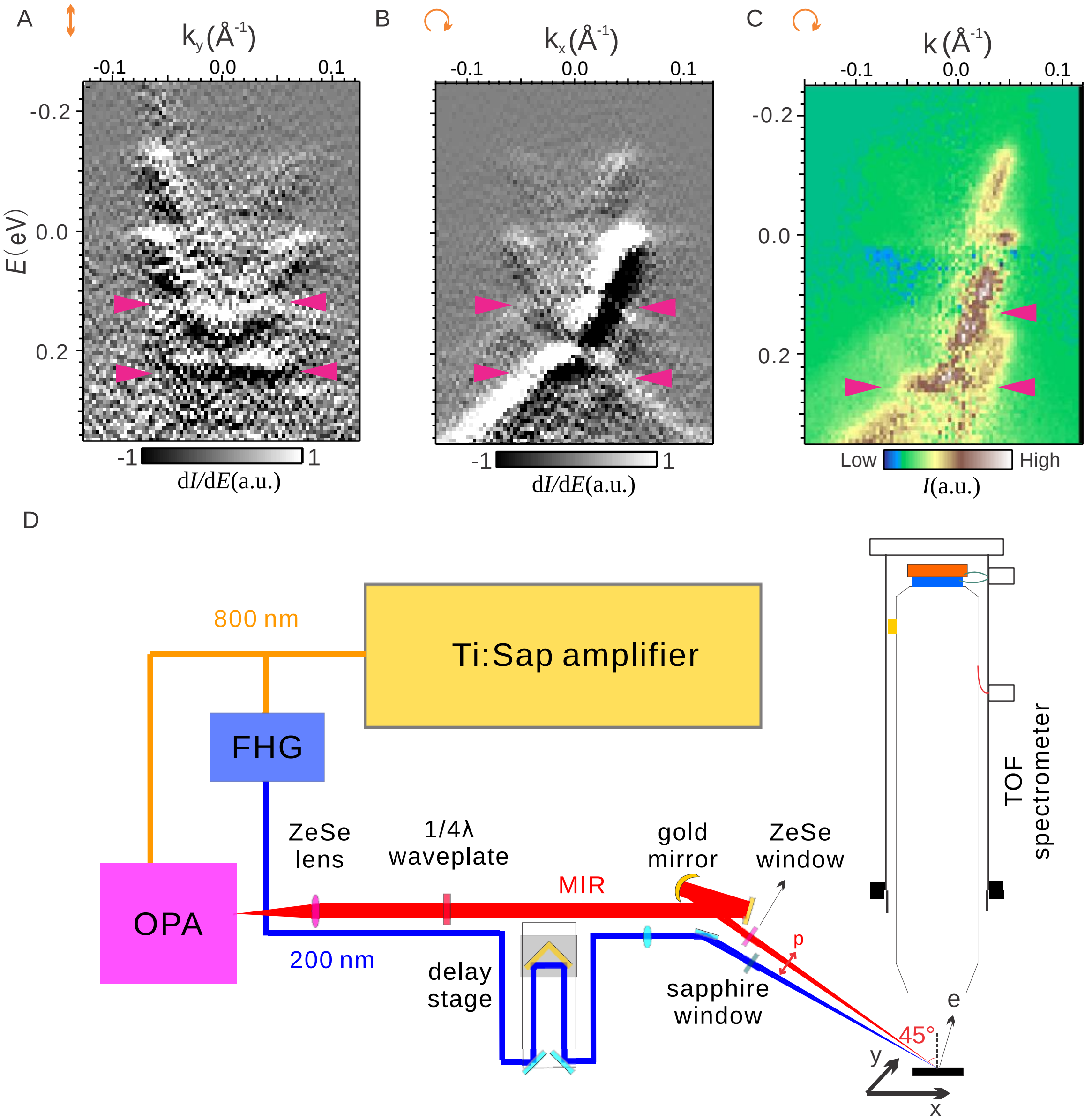

**Fig. S3**

Additional spectra showing band gaps at avoided crossings. **(A)** and **(B)** are derivative spectra under linearly and circularly polarized MIR light respectively. **(C)** Difference spectra along $45^0$ from $k_x$ under circularly polarized MIR pulse. The gaps are indicated by the pink triangles. **(D)** A sketch of the experimental setup.

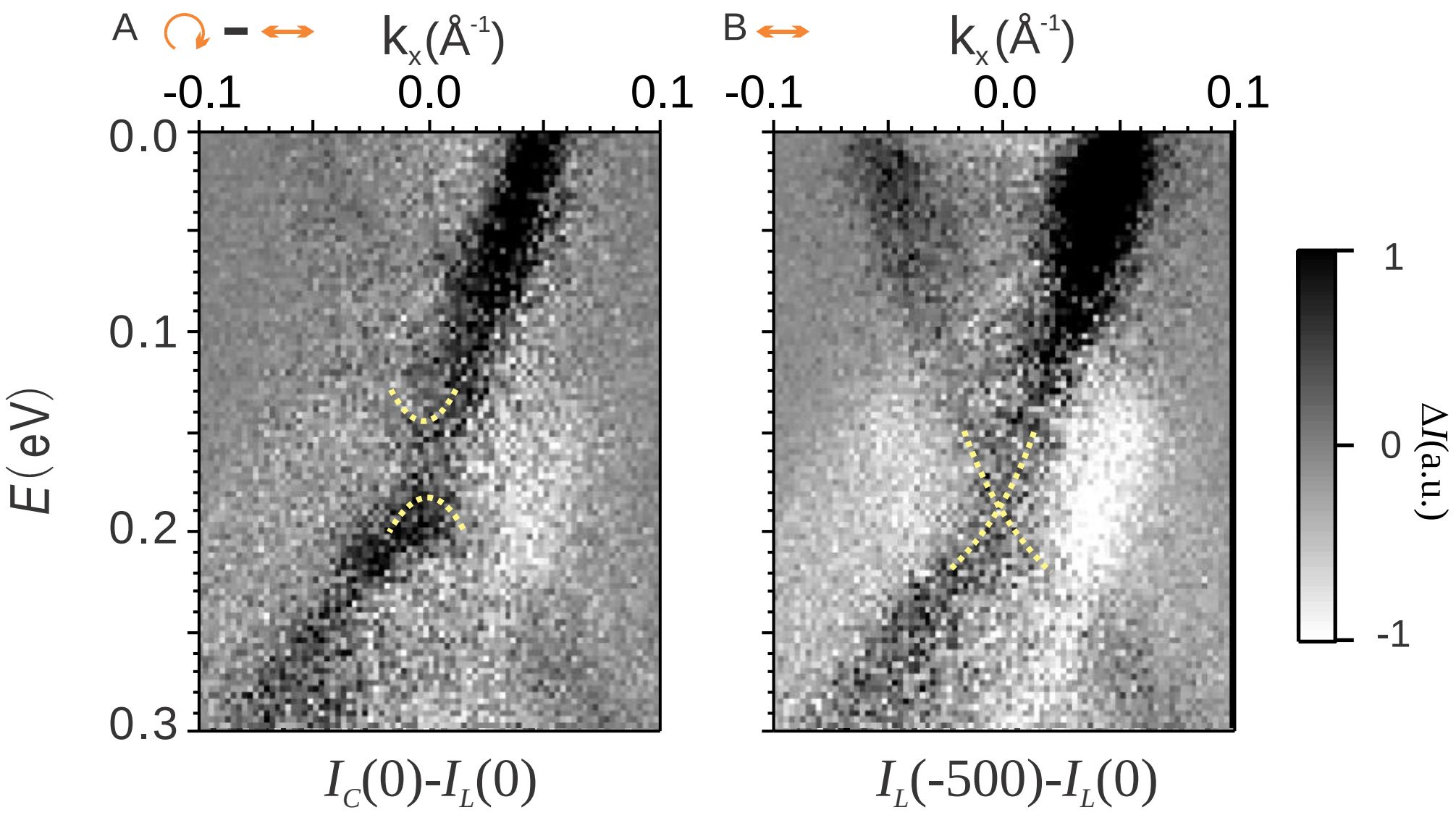

**Fig. S4**

Difference spectra along $k_x$. **(A)** is taken between spectra under circularly polarized light $I_C(t)$ and under linearly polarized light $I_L(t)$ at $t = 0$ fs. **(B)** is between spectra at $t = -500$ fs and $t = 0$ fs under linearly polarized light. Dashed lines are guides to the eye outlining the dispersion of the bands.